\documentstyle[preprint,prb,aps]{revtex}

\begin{document}
\draft

\title{Structural and magnetic properties of Fe/ZnSe(001) interfaces}

\author{B. Sanyal$^{\dagger}$ \cite{bs} and S. Mirbt \cite{sm}} 

\address{Department of Physics, Uppsala University, 
 Uppsala, Sweden} 
\address{$^{\dagger}$ present address : Maxlab, Lund University, Sweden}

\maketitle
\begin{abstract}
We have performed first principles electronic structure calculations to investigate the structural
and magnetic properties of Fe/ZnSe(001) interfaces. Calculations involving full geometry optimizations 
have been carried out for a broad range of 
thickness of Fe layers(0.5 monolayer to 10 monolayers) on top of a ZnSe(001) substrate. 
Both Zn and Se terminated interfaces have been explored.
Total energy calculations show that Se segregates
at the surface which is in agreement with recent experiments.  
 For both Zn and Se terminations, the interface Fe magnetic moments are higher than the bulk bcc Fe moment.
 We have also investigated the effect of adding Fe atoms on top of a reconstructed ZnSe surface to explore 
the role of reconstruction of semiconductor surfaces in determining properties of metal-semiconductor
interfaces. Fe breaks the
Se dimer bond formed for a Se-rich (2x1) reconstructed surface. 
Finally, we looked at the reverse growth i.e. growth of Zn and Se atoms on a bcc Fe(001) substrate
to investigate the properties of the second interface of a magnetotunnel junction. 
The results are in good agreement with the theoretical and experimental results, wherever available.

\end{abstract}
\vspace{20mm}

\narrowtext
\section{Introduction}
Spin electronics is one of the major topics of research nowadays where
one combines the charge and spin degrees of freedom of carriers \cite{prinz}. This has got
immense applications in modern technology e.g. magnetic recording industry and
semiconductor appliances. The recent exciting suggestions are to fabricate spin transistors 
and spin polarized tunneling junctions made up of ferromagnetic metals and semiconductors. 
 Large tunneling magnetoresistance (TMR) has been observed experimentally for heterostructures  
built of 
dilute magnetic semiconductors (DMS). There are also theoretical predictions of large TMR for
heterostructures based on Fe and semiconductors \cite{butler1}. In this
context, the interface between a metal and a semiconductor plays a crucial role in
determining the magnetic and hence the transport properties of the hybrid systems as the injection 
of a spin-polarized electron from a magnetic material to the semiconductor depends crucially on
the interface properties. The 
typical semiconductors used for these purposes are GaAs and ZnSe which belong to III-V and
II-VI compounds respectively. In such devices Fe is one of the most commonly 
used 3d metal as it has got a high Curie temperature. 

Informations about the structural and magnetic properties of the interface between Fe and ZnSe are available
from experiments \cite{jonker1,jonker2,bayreuther}. 
Jonker {\it et al.} studied Fe growth on ZnSe(001) at 450 K. 
The growth was smooth and epitaxial. However, they observed a reduction in magnetic moment 
of Fe compared to the bulk value.  
Recently, Reiger {\it et al.} \cite{bayreuther} 
studied Fe growth on ZnSe(001) at room temperature and
found a three dimensional nucleation at the beginning of the growth process followed by a coalescence 
of the 3D islands around 7 monolayers(ML). There was no reduction of the Fe magnetic moment even for 
a low Fe coverage. 

Theoretical investigations by ab-initio electronic structure calculations have been done
for Fe/ZnSe interfaces and multilayers \cite{butler1}, \cite{freeman}, \cite{dejonge}, \cite{dederichs}.
Continenza {\it et al.} \cite{freeman} studied various Fe/ZnSe superlattices by all-electron full potential
linearized augmented plane wave method within the local spin density approximation. They concluded that the 
enhanced Fe magnetism is suppressed as the Fe thickness is increased and the interface effects rapidly
die out in the inner Fe layers.
de Jonge {\it et al.} \cite{dejonge} have performed first principles electronic structure calculations of 
superlattices of Fe and several semiconductors (Ge, GaAs and ZnSe) by a localized
spherical wave method. 
Butler {\it et al.} \cite{butler1} studied the tunneling structures of ferromagnets and semiconductors by a 
layer Korringa-Kohn-Rostoker (LKKR) method. Moreover, they have investigated \cite{butler2} 
extensively the transport properties of magnetotunnel junctions of Fe and ZnSe using Landauer approach for
 the calculation of conductance.

All the above theoretical calculations were done for ideal interfaces i.e. the metal and semiconductor 
atoms followed a regular stacking.  
In this paper, we study the relaxed Fe/ZnSe interface, i.e we allow for
ionic, shape, and volume relaxations. We do not consider any temperature driven
reconstruction except segregation, because
it is known, that the semiconductor constitutents segregate \cite{segre} towards the surface when they are in contact with 
a metal. 
(We compared the total energies of different atomic configurations 
to investigate the surface segregation of Zn and Se.)
We study (section III) the dependence of the interface electronic properties 
on the Fe film thickness and the semiconductor termination.
Since, for a magnetotunnel structure, two interfaces are formed between
Fe and ZnSe, we also studied (section IV) the reverse growth i.e. Zn or
Se on a bcc Fe(001) substrate.
Moreover, we investigate (section V) the importance of
the semiconductor surface reconstruction for the interface formation.  

\section{Computational details}
Calculations have been performed by a self-consistent first-principles plane wave pseudopotential
code (VASP) \cite{vasp}. Vanderbilt type ultrasoft pseudopotentials \cite{uspseudo}
were used for the calculations which are reported to work 
well for transition metals \cite{hafnerfeconi}. More accurate PAW (projector augmented wave) \cite{paw} 
 pseudopotentials were also used 
 in some cases for testing of accuracy. Calculations reveal that the ultrasoft pseudopotentials 
are good enough to reproduce results with reasonable accuracies. An energy cut-off of 250 eV was
used for the kinetic energy of plane waves in the basis set. 
We have checked thoroughly that this cut-off energy is sufficient as it produces negligible Pulay
stress.
Perdew-Wang GGA(generalized gradient approximation) exchange -correlation \cite{pw} was
used instead of LSDA (local spin density approximation). LSDA description of the 
Fe ground state structure is inappropriate and the structural
informations of Fe layers obtained from geometry optimizations might be incorrect. 
We used a (1x1) unreconstructed surface cell with at least 10 \AA \ vacuum 
and six monolayers of ZnSe (three Zn and three Se layers). The number of Fe atoms varied from 1 (0.5 ML) to
20 (10 MLs) in the unit cell. The bulk layer at the bottom was terminated 
with two pseudo hydrogen atoms to avoid charge sloshing \cite{pseudo}. Two layers of ZnSe (one Zn and 
one Se) together with the pseudo hydrogen atoms were kept unrelaxed
at the ideal bulk positions. It is to be noted that the 
lattice constant of bcc Fe is almost half of that of ZnSe. This helps
 Fe to grow in a bcc phase on a ZnSe(001) substrate without 
appreciable strain. Our calculations with GGA yield equilibrium 
lattice constants of bcc Fe and zinc-blende ZnSe systems to be 2.86 \AA
\ and 5.75 \AA \ respectively.
The experimental lattice constant of ZnSe is 5.66 \AA \ and the overestimation of the calculated one is
obvious for using GGA.
Thus in experiment the Fe lattice constant is 1.4 \% smaller than the
ZnSe lattice constant, whereas in our calculation the Fe lattice
constant is 0.5 \% larger.
The calculated direct band gap at the $\Gamma$ point for the equilibrium lattice parameter is small (1.14 eV) 
compared to the experimental band gap (2.8 eV). As
a first step, we didn't attempt to correct that but this correction might be important from the
point of view of metal-semiconductor junction properties. 
 Total energies were always converged upto
10$^{-4}$ eV for electronic relaxations and 10$^{-3}$ eV for ionic relaxations. 
The atomic positions were relaxed alongwith the volume and shape of the unit cells. The full optimization with relaxations
 of internal and external degrees of freedom is important as far as magnetism is concerned. Forces were
converged with a tolerance of 10$^{-2}$ eV/\AA \ .  
The Monkhorst-Pack scheme \cite{monkhorst} was used for the generation of special k-points in the Brillouin zone. 
Convergence was checked carefully with respect to the 
number of k-points used. For example, a 6 $\times$ 6 $\times$ 2 mesh
for k points was used for the largest unit cell containing 20 Fe atoms 
which yielded 36 k points in the
irreducible Brillouin zone (IBZ) for the accurate calculation of DOS by 
tetrahedron method. For smaller cells,
we used 108 k points in IBZ. For the relaxation of the atoms, smaller set of k-points were
used (5 k-points in IBZ for a 20 atom cell) alongwith the Gaussian smearing of k-points. 
It should be mentioned here that there is no unambiguous way of 
determining the local properties e.g. local density of states, local magnetic 
moments etc. for a plane-wave method as the radius within which these properties are sought is 
not well defined. The plane wave components of the eigenstates were projected on linear 
combinations of spherical waves inside the atomic spheres around the atoms. The 
coefficients of the local spherical waves were used to construct the angular momentum
projected local DOS. Details have been discussed by Eichler {\it et al.} \cite{eichler}. 
For the present calculations, the radii of Fe, Zn and Se atoms were
chosen to be 1.41 \AA\, 1.35 \AA\ and 1.48 \AA\ respectively. These values were taken from the
 calculations of bulk bcc Fe and bulk ZnSe systems.

\section{Different Fe coverages}
\subsection{0.5 ML coverage}

We first present results for the calculations involving ideal stacking positions.
Though this situation is not energetically favourable compared to the relaxed case, we investigate this as a 
starting point.
Total energies are compared for the two cases 
(i) Fe is on the top of the ZnSe substrate in a regular stacking position 
(ii) Fe is in a substitutional
Zn (Se) site in a subsurface layer in exchange with a Zn (Se) atom kept on the surface. 
We have found that Fe in a substitutional Zn site in the subsurface layer is more favourable
than Fe being on the surface. The energy difference between the two configurations is 0.83 eV/cell. 
But, for Se substitution, Fe 
 prefers to sit in the surface keeping the Se atom in the subsurface layer. Here, 
the energy difference is 0.40 eV/cell. However, 
these cases without relaxations are unfavourable compared to the relaxed case.
 It is well known that in the case of dilute magnetic semiconductors (DMS) of (Fe,Zn)Se \cite{dms2}, 
Fe prefers to sit in the cation site.   
For this case, the calculated Fe-Se distance (2.47 \AA \ ) is 
slightly smaller than the Zn-Se bondlength (2.49 \AA \ ), whereas the
measured Fe-Se distance (2.48 \AA \ ) is slightly larger than the measured Zn-Se
bondlength (2.45 \AA \ ). The magnetism is 
atomic like with an integer magnetic moment of 4.00 $\mu_{B}$/cell. 

Now we discuss the results of the relaxation studies.
The structure of the region very close to the interface of a metal and a semiconductor  
is an important factor for spin injection from a metal to a semiconductor. Since semiconductors
have open structures, the metal atom will proceed inside the semiconducting substrate, where the charge overlap 
with the substrate atoms is increased. 
To investigate this possibility, an approximate energy landscape 
has been calculated by putting half a monolayer of Fe in different 
positions on and inside the host ZnSe lattice and comparing the total energies. 
The result is shown in Fig.1 for a Se-terminated surface. 
The energy difference is defined as $\Delta E = E^{B}_{tot}
-E^{S}_{tot}$ where $E^{S}_{tot}$ is the total energy for an Fe atom sitting in the regular position on top of the surface 
and $E^{B}_{tot}$ is the total energy 
of the Fe atom sitting on different positions within the substrate.  
The most preferable position we found 
to be an interstitial one which is 2.875 \AA \ below the surface. 
The corresponding total magnetic moments of the unit cells are 
plotted in the inset. 

The final atomic configurations of the most preferable positions are shown in Fig. 2(a-c) and 
3(a-c) for the Zn and Se terminated cases respectively. 
One can note the difference in the relaxations in the two cases: 
For a Zn-terminated case (shown in Fig. 2), with relaxations of atoms, the surface Zn atom moves 
upwards leaving a vacancy which is eventually occupied by the 
burried Fe. The topmost Zn atom occupies an ideal
stacking position. This also conforms with the unrelaxed calculation mentioned above.

For a Se terminated case (shown in Fig. 3), 
starting from an identical situation as before, Se moves a little bit upwards.   
The burried Fe atom remains more or less in the interstitial position.  
So, the distinct difference in the two cases is the site
occupancy of Fe. In both cases, relaxation leads to lower
total energies compared to the cases of unrelaxed regular stacking as discussed before (0.03 eV/cell and 0.8 eV/cell 
lower for Zn and Se terminated cases respectively). 
It is seen that for Zn termination, the nearest Fe-Zn
and Fe-Se distances are 2.49 \AA \ and 2.37 \AA \ respectively. For Se coverage, the distances are
2.96 \AA \ and 2.52 \AA \ respectively. 

(So, magnetic moment is enhanced compared to that of bulk bcc Fe.   
However, it should be kept in mind that as we are using GGA, calculated value of magnetic moment 
 (2.33 $\mu_{B}$) is greater than the experimental value of 2.2 $\mu_{B}$ for bulk bcc Fe.) 

Atom and orbital projected DOSs are shown in Fig. 2(d-h) and Fig. 3(d-h) 
for Zn and Se terminated cases respectively.
For both terminations, the majority Fe DOS has no states
at the Fermi energy  
and we find an atomic like integer moment of 3.00 $\mu_{B}$ for the unit cell. 
 This is not found for cases without relaxations described before. 
The magnetic moments for the unrelaxed cases are greater and of
non-integer type. Half metallicity is seen only when relaxation is allowed. 
Half metallic systems (a system where one finds a
metallic DOS for one spin channel and semiconducting DOS for the 
other showing 100 $\%$ spin polarization at the Fermi
energy) are extremely important in the context 
of magnetoelectronics involving spin transport between ferromagnetic
systems across a semiconducting/insulating barrier. 
However, with increased Fe coverage, this behavior dies out and we
arrive at a normal metallic DOS.  
 A similar calculation for Co for 0.5 ML coverage also yields an integer moment of
2.00 $\mu_{B}$/cell with both Zn and Se on top of burried Co. 

Comparing this case with the total magnetic moment of 4.00 $\mu_{B}$ of a 
bulk DMS (Fe,Zn)Se, it is clear that the difference is 
due to the occupancy of the minority spin states of Fe as the majority
states are completely occupied for the two cases.
In DMSs, Hund's rule for the atomic moment is followed 
resulting in 
the occupation of 5 and 1 electrons in the majority 
and minority spin channels of Fe $d$-states respectively. 
In the present case, 
again Hund's rule for the atomic moment is followed, but Fe has gained one electron, such that now two electrons are in the minority
spin channel. The reason is the following: At the ideal surface, i.e
without Fe, the two Zn (Se) dangling bonds (DB)
are each partially occupied with 0.5 (1.5) electrons. A semiconductor
gains energy by having either occupied or unoccupied DB.
Usually, anion DB (Se) are fully occupied and cation (Zn) DB are empty. 
Therefore, taking one electron from the Zn (Se) DB will lead to either
occupied or unoccupied DB: for the Zn termination two empty DB and for
the Se termination one empty and one occupied DB. This electron from the DB is transferred to Fe and gives rise to the two electrons
in the d-minority channel.  

Now, let us look at the projected DOS in detail. 
In both sets of figures (fig. 2(d-h) and fig. 3(d-h)), 
Fe {\it s} and {\it d} states,
Zn {\it s}, {\it p} and {\it d} states and Se {\it s} and {\it p} 
states are shown (see figure captions for details).  
The energy scale is chosen to avoid showing 
the huge peak for Zn 3{\it d} state which is deep in energy and doesn't 
hybridize with the Fe {\it d} states.
For both terminations, the Fe DOS looks quite 
different from that of bcc bulk Fe. The majority spin states 
are filled up for Fe. 
For Zn termination, in Figs. 2(e) and 2(g), projected DOSs for surface(Zn$_{1}$) 
and sub-surface Zn(Zn$_{2}$) atoms are shown. 
The relative distribution of {\it s} and {\it p} states 
are different with the subsurface Zn (Fig.2(g)) having more bulk like character.
Fe {\it d} states are narrower compared to the same for Se-terminated 
case (see Fig. 3(d)) due to less number of
nearest neighbours in a substitutional position 
compared to an interstitial position for Se terminated case. 
This yields a slightly larger local magnetic moment of Fe for the Se terminated case (Table I). 
For Se termination, the effect of the Fe atom sitting in an 
interstitial position is pronounced even in the 
states of the atoms of deep subsurface layers e.g. Se$_{2}$ and Zn$_{2}$.
 The sharp peaks in the conduction bands have mostly
Zn 4{\it s} and Se 4{\it s} and 4{\it p} characters. 
The induced spin polarizations of Zn and Se atoms at the surface are evident from the
relative shift of majority and minority states (Table I). 

\subsection{1.5 ML coverage}
This case corresponds to 3 Fe atoms in the unit cell. 
As a complete Fe monolayer should contain 2 Fe atoms and already 
1 Fe atom is burried in a subsurface layer, 
it is interesting to find the configuration with the segregated Se atom.
For a Se terminated cases, we explored three different situations 
to find the lowest energy configuration
(see Fig. 4). In Fig. 4(a), Se is sitting at a subsurface layer covered by a complete Fe layer
at the surface. In Fig. 4(b), the surface is covered by 0.5 ML of Fe and Se is still in a 
subsurface layer. Finally, in Fig. 4(c), Se is at the surface followed by a complete Fe monolayer.  
The resulting sequence of energetics is E$_{c}$ $<$ E$_{b}$ $<$ E$_{a}$. The energy
difference between configuration (c) and (b) is quite large (2.6 eV/cell). 
So, in summary, Se segregates
towards the surface. In all cases, the magnetism is almost similar. In the most favourable
case (c), the two Fe atoms (indexed 1 and 2 in the figure) 
below the surface have magnetic moments of 
2.84 $\mu_{B}$ and 2.66 $\mu_{B}$ respectively. The burried Fe
atom (indexed as 3) has a magnetic moment of 2.92 $\mu_{B}$. 
The surface Se atom (indexed as 4) attains an
induced magnetic moment of 0.07 $\mu_{B}$. Fe(1)-Se(4) distance is 2.48 \AA\ which is different than 
the cases where Fe and Se lie almost in the same layer.  
Though we have performed calculations for different thicknesses of Fe layers, we present only the details
of the important ones. The next section describes a case with a relatively thick Fe layer. 

\subsection{20 atoms (10 MLs) of Fe}
To see the effect of the interface on the bulk Fe atom like layers, 
we performed calculations with 20 atoms 
of Fe on ZnSe substrate. We explored both the cases of Se-segregated 
and non segregated cases. The configuration
with Se at the surface has the lower energy (energy difference between the surface segregated and
non-segregated profiles is 1.5 eV/cell). For a Se-segregated case, 
the surface Fe-Se distance is 2.81 \AA\ . Magnetic moment of the 
surface Fe atom is 2.96 $\mu_{B}$. The burried Fe atom in the interface 
has a moment of 
2.88 $\mu_{B}$ with a Fe-Zn distance of 2.87 \AA\ . Informations about the structural and magnetic properties
are listed in table II for a few selected cases. 

In Fig. 5, we have shown the layer projected local magnetic moments for the case of 20 Fe atoms deposited
on ZnSe. The surface and the interface Fe atoms have large magnetic moments compared to the bulk value.
But the atoms attain the magnetic moments of bulk bcc Fe very near to the interface. 
The induced moments on Se and Zn atoms close to Fe are parallel and antiparallel with respect to Fe moments.
The corresponding
projected DOSs are shown in Fig. 6. 
 Figs. 6(a) and 6(b) show the surface Fe and Se atoms respectively. Fig. 6(c) shows the PDOS for
an Fe atom in an intermediate Fe monolayer and (d) shows the same approaching the interface. 
Figs. 6(e) and (f) are PDOSs of 
burried interface Fe atom and Zn atom respectively. For the surface atom, the majority DOS is full whereas the 
minority
DOS is partly occupied. Both the bonding and the antibonding {\it d}
 states of majority spin are filled whereas the bonding 
states for minority spin are occupied leaving the antibonding state fully empty. 
Fig. 6(c) corresponds to the bulk bcc Fe DOS where the majority states are occupied (both bonding and the 
antibonding states are filled) and the 
Fermi level lies at the pseudogap of the minority DOS separating bonding and antibonding {\it d} states. 
The majority DOS is more affected than the minority one while going from surface to the interface. 
Approaching the interface, the exchange splitting between the
majority and minority DOSs increase and for the interface atom which is marked as Fe$_{I}$, 
the minority DOS is pushed towards the low energy side. DOS looks similar to that obtained
 by other authors \cite{dejonge}. DOS at
Fermi level is considerably different for different Fe atoms. The spin polarization SP is defined as 
SP=$n(E_{F})_{\uparrow}-n(E_{F})_{\downarrow}$/$n(E_{F})_{\uparrow}+n(E_{F})_{\downarrow}$ 
where $n(E_{F})_{\sigma}$ is the 
DOS at Fermi level for spin $\sigma=\uparrow,\downarrow$. 
SP is different for different Fe atoms. It is increased for the
surface and the interface Fe atoms (negative) 
whereas it has got a lower value (positive) for the bulk like Fe atom. 

In Fig. 7, we show the total and averaged magnetic moments of Fe 
atoms per unit cell as a function of Fe thickness.
The values of the total magnetic moment increases 
linearly with Fe thickness. This is also seen in experiments 
\cite{bayreuther}. The value of the slope is close to the 
bulk Fe moment but still larger because of the presence of
Fe surface atoms. For Fe/GaAs(001) interface, it is seen from experiments \cite{fegaas} 
that the extrapolated curve in a magnetic moment vs. thickness plot touches the 
thickness axis above the zero value. 
This means that for low thickness, magnetism of the interface is reduced
compared to the bulk value.
The same for Fe/ZnSe(001) interface doesn't show any reduction of magnetic moment 
at the interface. The reason is that the p-d wavefunction overlap between Fe and As is stronger than between Fe and Se. For example,
the Fe magnetic moment in an As environment is considerably reduced, whereas it is atomic like in a Se environemt.  
The inset shows the averaged magnetic moment per Fe atom in the cell as a function of Fe thickness.
Averaged moment is defined as the ratio of the total magnetic moment and the
number of Fe atoms in the unit cell.
It is seen that this value decreases with thicker Fe layer and 
approaches the bulk magnetic moment of Fe. This is 
understandable because as we deposit more and more Fe atoms, 
the effect of surface and interface washes out and the bulk
character is achieved.  

The DOSs at the Fermi levels for both Se and Zn terminated  
 cases are depicted in Fig. 8. It is to be noted that for these 
cases, surface segregation of Se and Zn atoms have also been considered. For that reason, a Zn atom is
close to Fe for a Se-terminated case and a Se atom is close to Fe for a Zn-terminated case. 
A segregated profile for a Zn-terminated case has been chosen intentionally to have a comparative study
with the Se-terminated case though Zn segregation has not been confirmed by the experiments.
The sequence of the atoms from left to right 
is from the interface towards the bulk semiconductor. 
For 1 Fe atom, the spin polarization(SP)
is -1.0(i.e. 100 $\%$ spin polarization). For 4 and 20 Fe atoms, the spin polarization is reduced. 
For the thickest layer i.e. 20 atoms of Fe, we find that
the spin polarization reduces faster to zero as one approaches the bulk layers. 
For Se terminated cases (figs. 8(a-c)), spin polarization is always negative but for 
Zn terminated cases, the same for 4 and 20 atoms Fe (figs. 8(e-f)) shows positive values.

\section{Reverse growth}

To prepare a magnetotunnel junction, one has to care for two interfaces. One where the semiconductor atoms
are deposited on the metal electrode and the other where the metal electrode atoms are deposited on 
semiconductor substrate. It has been observed by transmission electron microscopy \cite{fred_thes} 
that the interface structures are not the same for
the two cases. The upper interface (Fe/ZnSe) is rougher than the lower one (ZnSe/Fe). 
To investigate this, we have done very simple
calculations where we put either Zn or Se atoms on a bcc Fe(001) substrate. 
We compared the total energies of two configurations
(i) Zn or Se atom is on top of Fe substrate 
(ii) Zn or Se atom is in a subsurface layer covered by one Fe layer at top.
We have found that the semiconductor 
constituents prefer to sit on top of a Fe substrate indicating a sharp interface.
For Se, the energy difference is 2.15 eV/cell
 whereas for Zn, it is 0.9 eV/cell. It suggests that Se/Fe interface is less
 probable to be a mixed one compared to the Zn/Fe interface. 
However, the magnetic moments are not changed
appreciably for different configurations.
 We have already reported in the previous 
sections that for Fe/ZnSe interface, half a monolayer of Fe
prefers to stay in a burried position in the 
ZnSe host. We can conclude that this interface is less sharp than 
the one with the reverse growth. 

\section{Fe on a reconstructed surface}
In this section, we present the results of 
depositing Fe atoms on a Se-rich (2x1) reconstructed surface. It is already
known from experiments \cite{recon_exp} and theoretical calculations \cite{chadi} 
that the Zn-rich and Se-rich surfaces undergo c(2x2) and (2x1)
surface reconstructions respectively. For a Se-rich (2x1) reconstructed surface, 
a Se dimer of bond-length 
2.40 \AA\ is formed on the surface \cite{chadi}. 
We have put 2 Fe atoms on top of a Se-rich (2x1) reconstructed surface
 having a Se dimer (Se1 and Se2) and started relaxing the atoms. 
In the final relaxed configuration, 1 Fe
atom sits in the same layer as one of the Se atoms(Se1) of the dimer breaking the Se dimer bond. 
The distance between Se1 and Se2 atoms becomes 5.24 \AA\ .
As the Se dimer is broken, the effect of the reconstructed semiconductor substrate is diminished.
The other Se atom(Se2) goes beneath
the layer. Another Fe atom goes to the surface. 
This is in accordance with all the cases where 1 Fe atom wants 
to be burried under the Zn or Se layer. 
The bond length between two Fe atoms and Se1 is 2.50 \AA\ each.     
The magnetic moment is increased compared to the unreconstructed 
case by 0.5 $\mu_{B}$/cell. This increase is due to
the lack of nearest neighbours of the topmost Fe atom. 
As the signature of the reconstructed semiconductor substrate
is lost, we therefore conclude that the surface reconstruction has no dramatic
effect in determining the properties of the Fe-ZnSe interfaces.

\section{Conclusion}

In this paper, we have extensively studied structural and magnetic properties of Fe/ZnSe(001)
interfaces. Geometry optimizations have provided useful structural informations of the interface. 
Our results show that half a monolayer of Fe prefers to sit below the surface.
For the Se terminated case, Fe prefers to sit in the interstitial site whereas for Zn terminated
case, Fe prefers a Zn substitutional site. We have observed surface segregation of 
Zn and Se atoms. However, the magnetic properties are not 
changed appreciably for segregated and non-segregated
profiles. 
Magnetic moment of the interface Fe atom is 
increased compared to the bulk bcc Fe moment
 for both Zn and Se terminations. Fe attains its bulk moment very close to the interface. 
A reverse growth i.e. Zn or Se on top of the bcc Fe surface has also been studied.
For a
Se-rich (2x1) reconstructed surface, Fe breaks the Se dimer bond and sits in between to have a 
proper bonding configuration. Effect of the reconstructed semiconductor substrate is diminished.

\acknowledgements
S.M. is grateful to the Swedish Natural Science Research Council and G\"{o}ran Gustafsson Foundation for 
financial support.

\begin{figure}
\caption[fig1] {Energy difference plotted as a function of the distance (in \AA\ ) of the Fe atom from the
 Se-terminated surface. See text for the details.}
\end{figure}

\begin{figure}

\caption[fig2] {Final relaxed configuration (a-c) from different perspectives 
and corresponding atom and orbital projected DOSs for 0.5 ML Fe on a
Zn-terminated surface. For figures (a-c), open circles, filled circles and open squares denote Zn, Fe
and Se atoms respectively. Vac. represents vacuum and the two vertical dashed lines denote the basic
(1x1) unit cell. Lines are drawn between atoms to show the bonds of the host zinc-blende
 lattice. For figures (d-h),
besides Fe, Zn and Se atoms are indexed according to the stacking in unit cell 
from surface towards the bulk. Appropriate legends are provided for the orbitals in the figures. The vertical
lines in figures (d-h) represent the Fermi level.}    
\end{figure}

\begin{figure}
\caption[fig3]{Same as Fig. 1 but for a Se-terminated case.}
\end{figure}

\begin{figure}
\caption[fig4] {Final relaxed configurations for 1.5 ML of Fe (3 Fe atoms in the unit cell) 
on a Se-terminated surface.
(a) A complete ML of Fe at the surface 
and Se is at the interface (b) Se has gone towards the surface (c) Se covers the surface followed by Fe
atoms. The other notations are similar to those of Fig.1 and Fig.2.}
\end{figure}
 
\begin{figure}
\caption[fig5] {Layer projected magnetic moments for the case of 20 Fe atoms on Se-terminated (and segregated)
surface. Note that for each Fe monolayer, there are two Fe atoms except for the surface and interface. 
The bold line indicates the zero value and the dashed line indicates the calculated magnetic moment
for bcc bulk Fe within GGA.}
\end{figure}

\begin{figure}
\caption[fig6]{Selected atom and orbital 
projected spin-resolved DOS for the case described in the previous figure. 
The sequence of the figures is 
 (a) surface Fe atom (Fe$_{S}$)(b) segregated Se atom (Se$_{S}$)(c) Fe at some intermediate layer (Fe$_{B}$)
(d) Fe approaching the interface(Fe$_{B-I}$) (e) Fe at the interface(Fe$_{I}$) 
(f) Zn close to Fe at the interface(Zn$_{I}$). 
The vertical lines show the position of the
Fermi energy.}
\end{figure}

\begin{figure}
\caption[fig7]{Evolution of magnetic moment with increasing number of Fe atoms in the unit cell. 
The graph is for Se terminated and
segregated case. The bold line is the fitted one with the calculated data (dashed line). 
(Inset) Average 
Fe moment/unit cell  vs. number of Fe atoms.} 
\end{figure}

\begin{figure}
\caption[fig8]{Densities of states at Fermi level $n_(E_{F})$ for (left) Se terminated and segregated
and (right) Zn terminated and segregated cases. (a) and (d) are for 1 Fe atom, (b) and (e) are for 4 Fe atoms
and (c) and (f) are for 20 Fe atoms. The upper triangle plots are for the 
majority spin electrons whereas the lower
triangles are for the minority spin electrons. 
The sequence of the atoms is from the interface towards the bulk
semiconductor (left to the right of the figures).} 
\end{figure}

\newpage
\onecolumn
\vspace{15mm}

\noindent{\bf TABLE I.} Magnetic moments (in $\mu_{B}$) for 0.5 ML of Fe.
\vspace{15mm}

\begin{tabular}{|l|l|l|l|l|}
\hline \hline 
                         & Fe moment & surface Zn moment & surface Se moment & total moment/cell \\
\hline
Zn-terminated            & 2.93      & -0.023            &                   & 3.00 \\         
\hline
Se-terminated          
			 & 2.84      &                   & 0.08              & 3.00  \\
\hline \hline
\end{tabular}

\vspace{15mm}

\noindent{\bf TABLE II.} Structural and magnetic informations for selected thicknesses of 
Fe layers. All these results are for Se-terminated and segregated cases. The subscripts
$S$ and $I$ denote surface and interface atoms. Fe$_{I}$ is in the burried position
close to the interface Zn atom. The values in the parentheses ()$^{\dagger}$ and 
()$^{*}$ are for Fe$_{S}$-Zn$_{S}$ and Fe$_{I}$-Se$_{I}$ distances for a Zn-terminated case
 with the corresponding local moment shown in ().

\vspace{15mm}

\begin{tabular}{|l|l|l|l|l|}
\hline \hline 
Thickness of Fe                   & Fe$_{S}$-Se$_{S}$ & Fe$_{I}$-Zn$_{I}$ & Fe$_{I}$           \\
layers                            & distance ( \AA\ ) & distance ( \AA\ ) & moment ($\mu_{B}$) \\
\hline                                                                                   
0.5 ML                            &  2.52 (2.49)$^{\dagger}$  & 2.96 (2.37)$^{*}$ &  2.84 (2.93) \\
1.5 ML                            &  2.48                     & 2.78              &  2.89        \\
2.0 ML                            &  2.74                     & 2.77              &  2.74        \\
10.0 ML                           &  2.81                     & 2.87              &  2.88        \\
\hline \hline
\end{tabular}

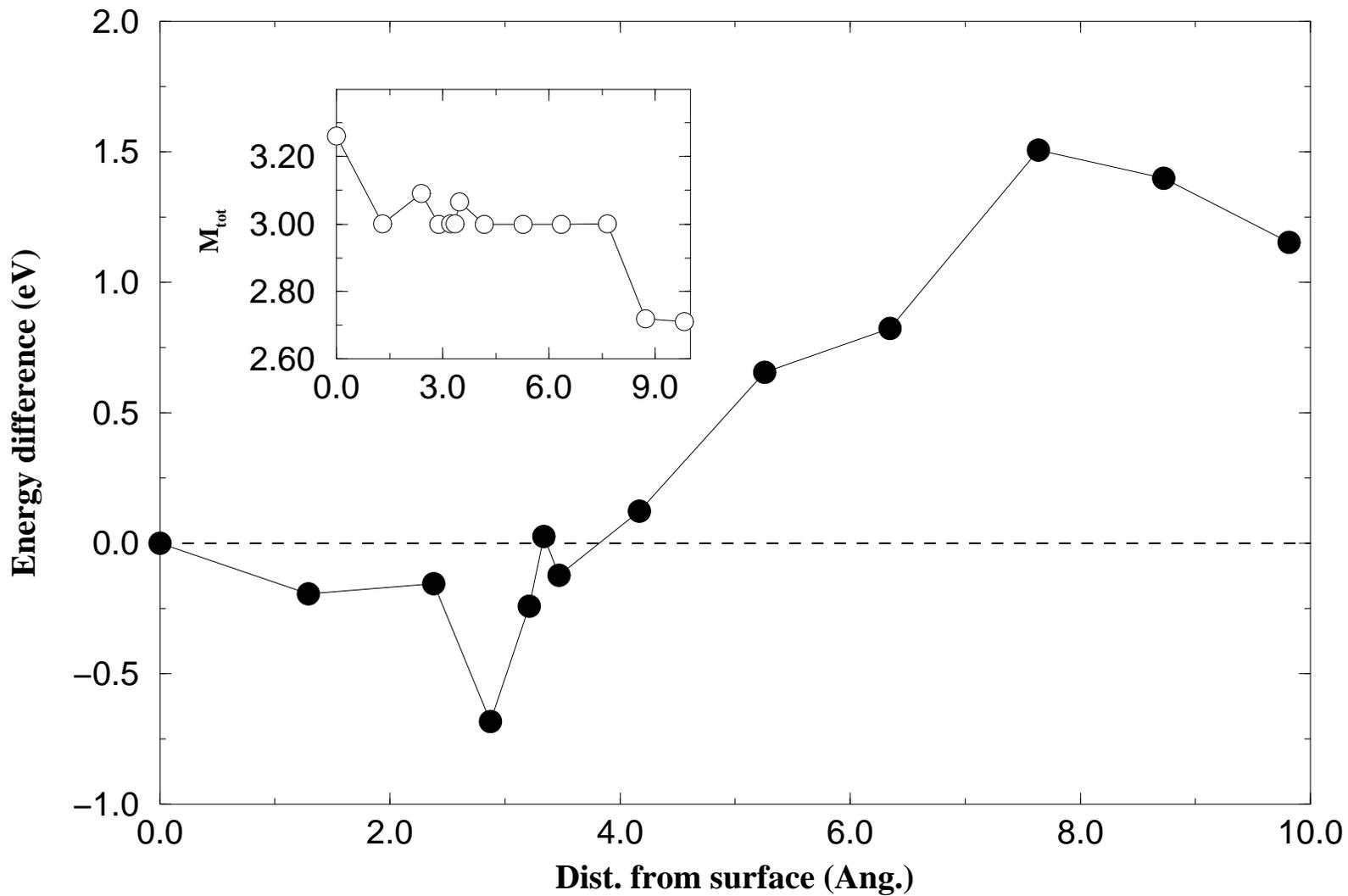
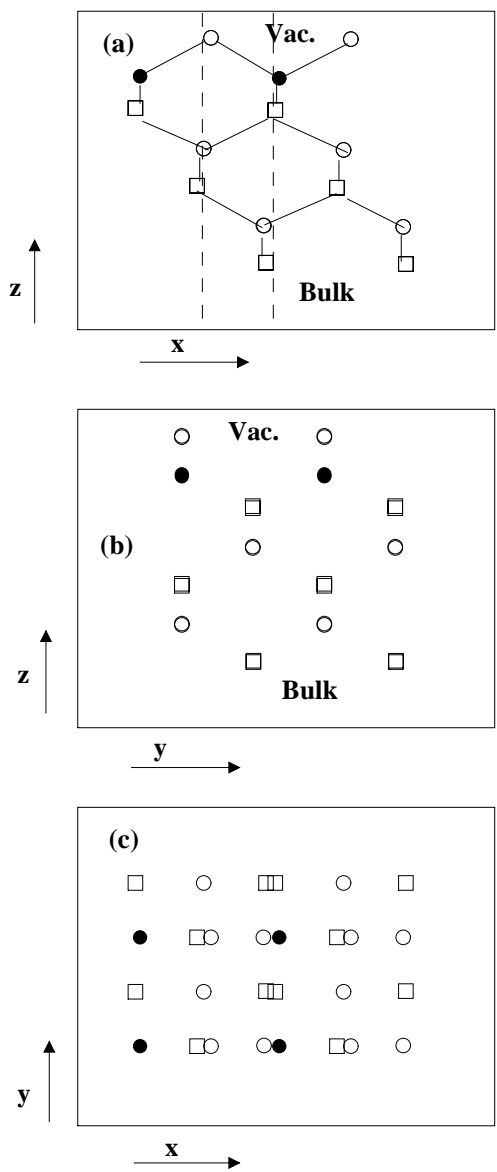
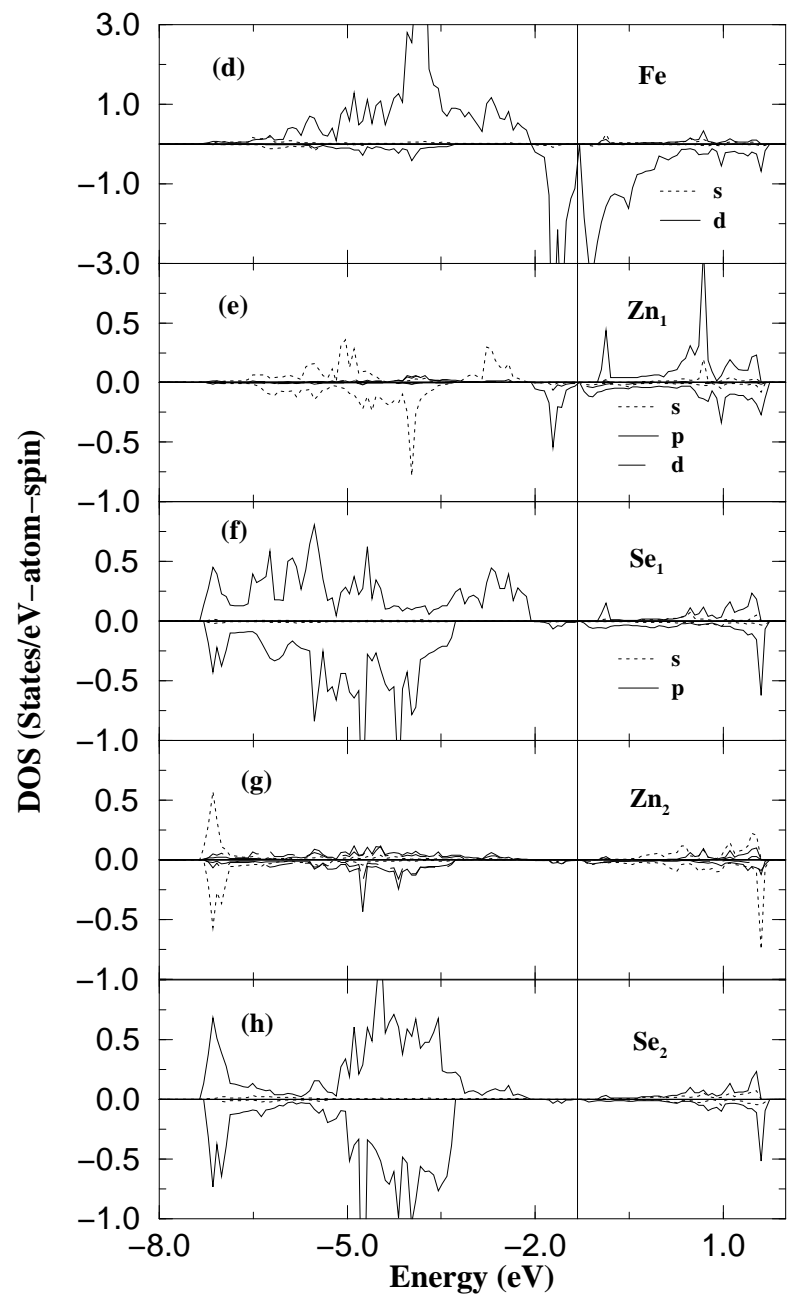
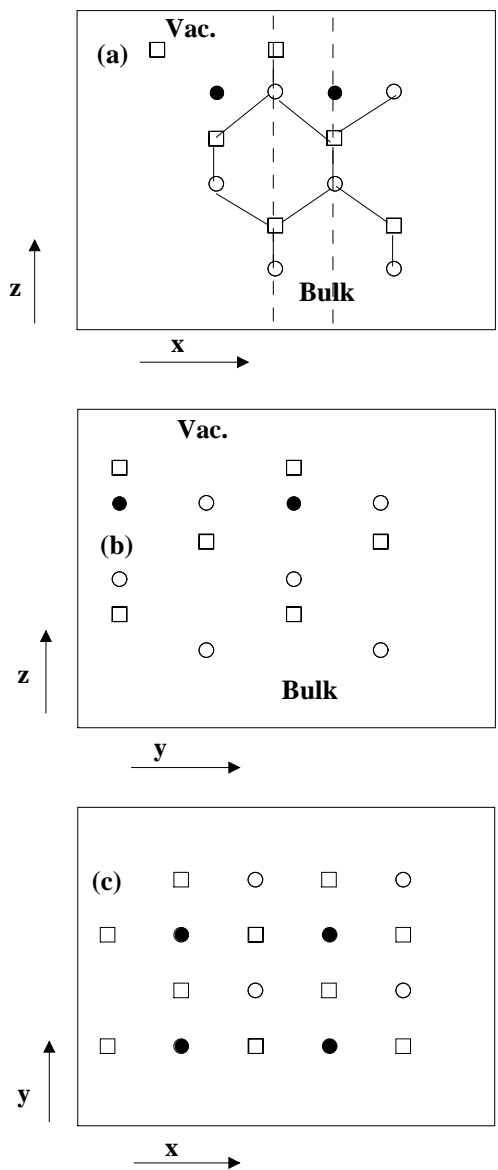
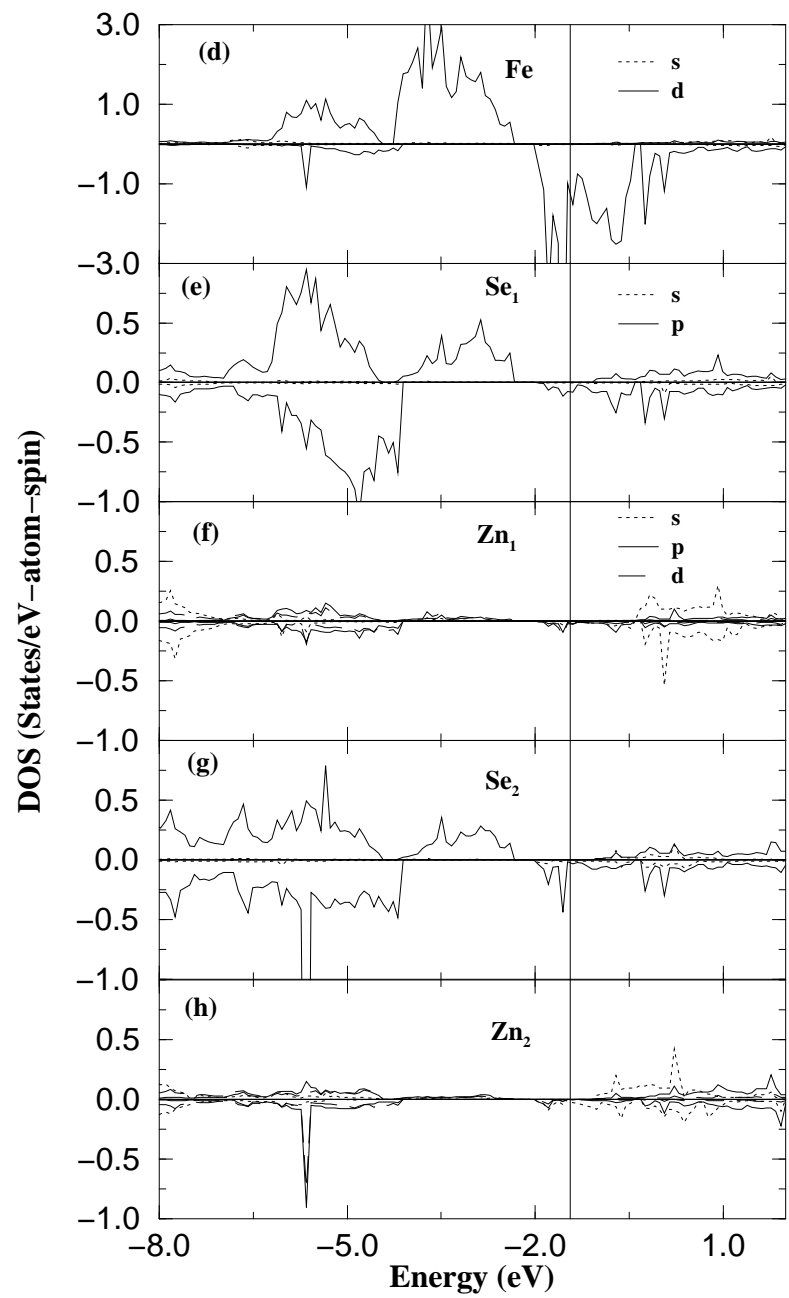
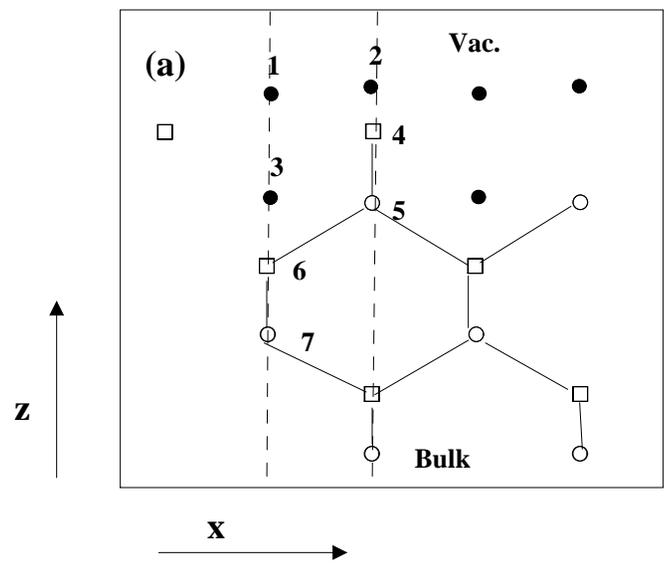
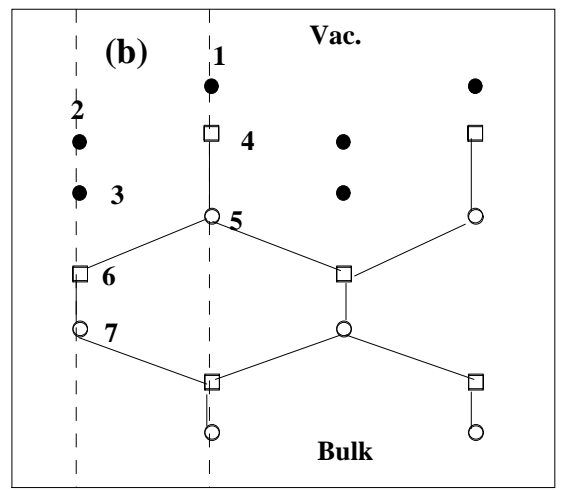
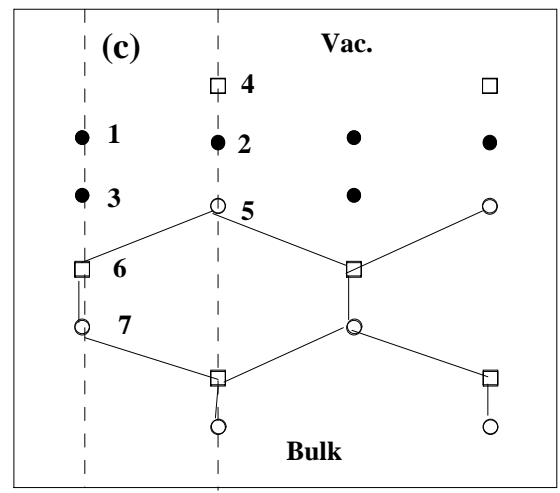
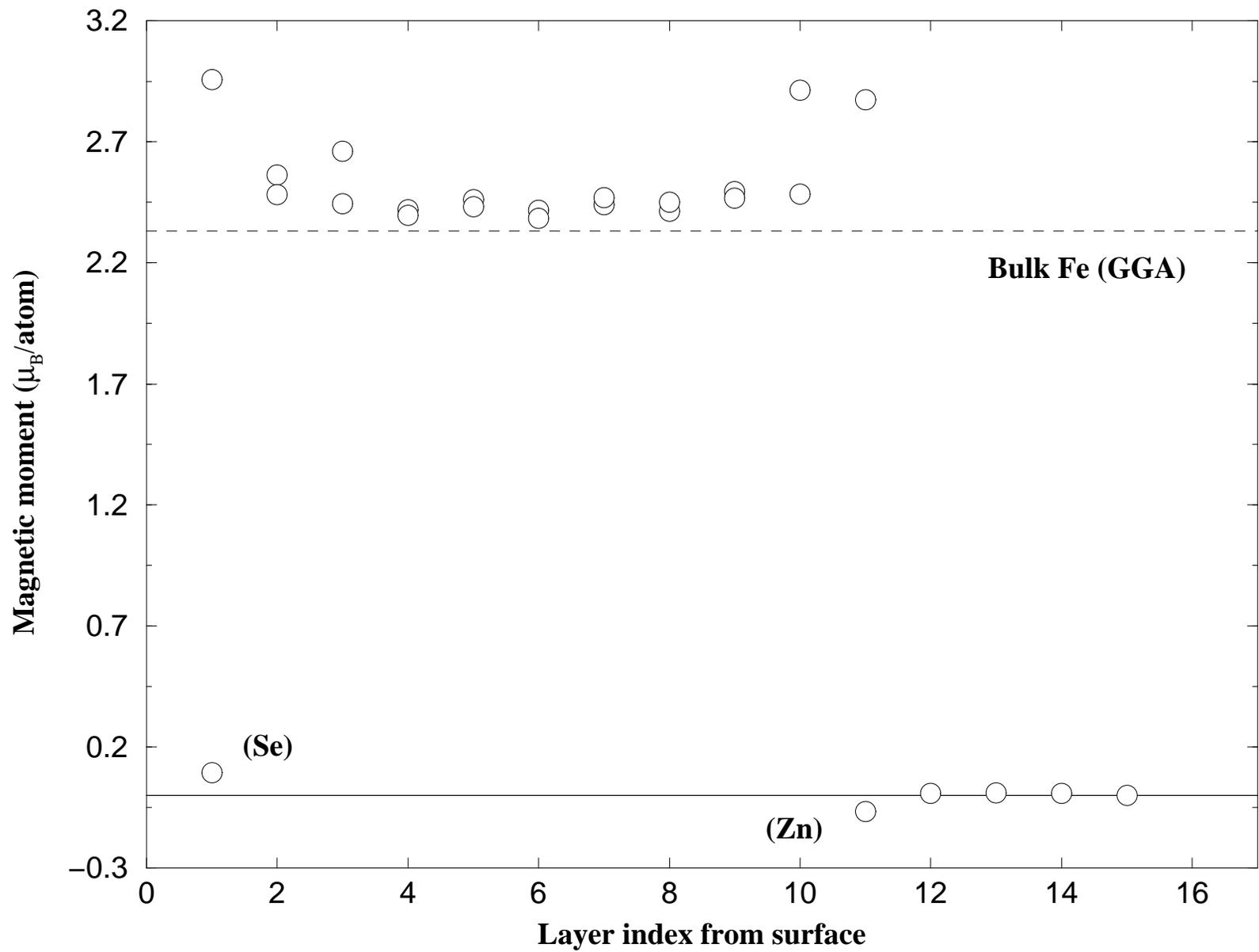
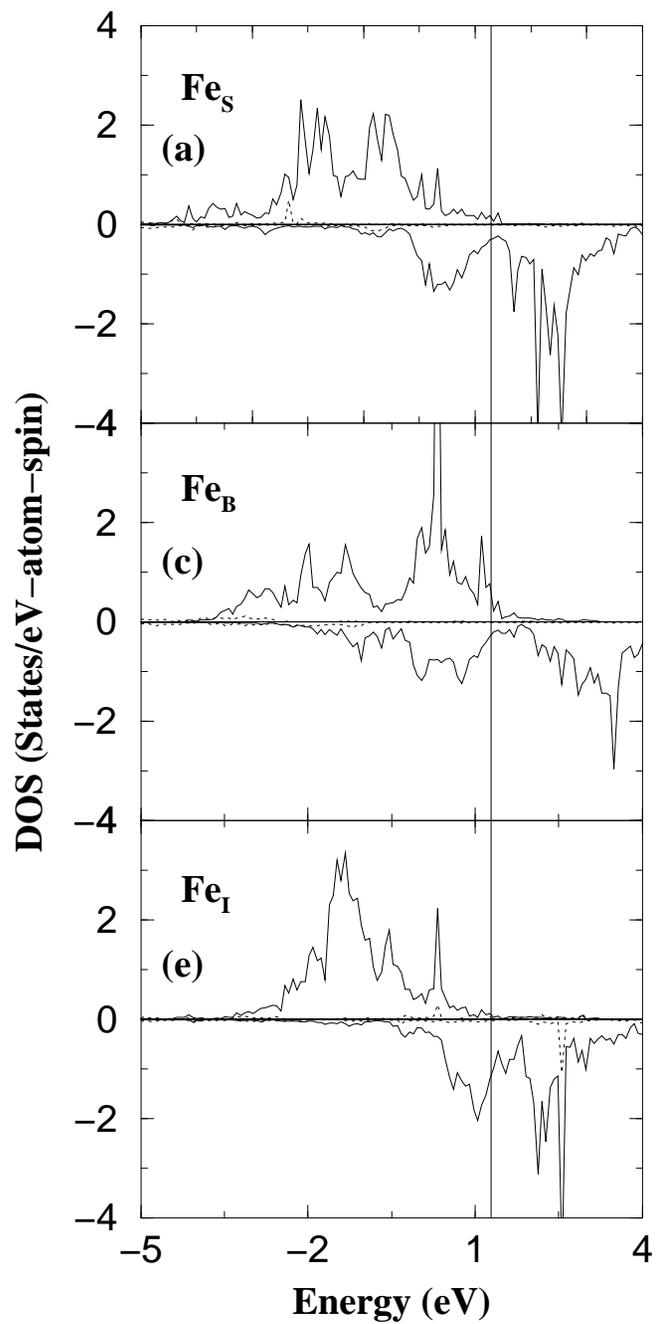
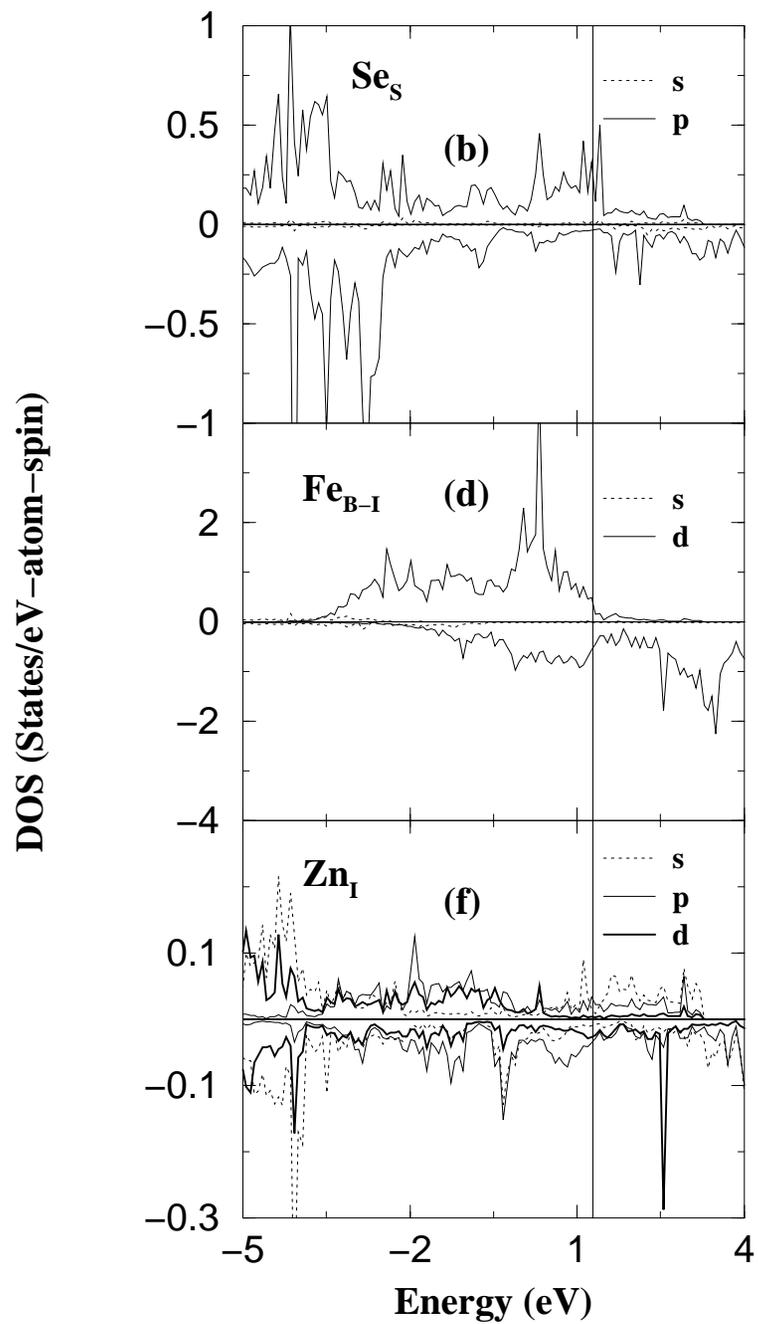
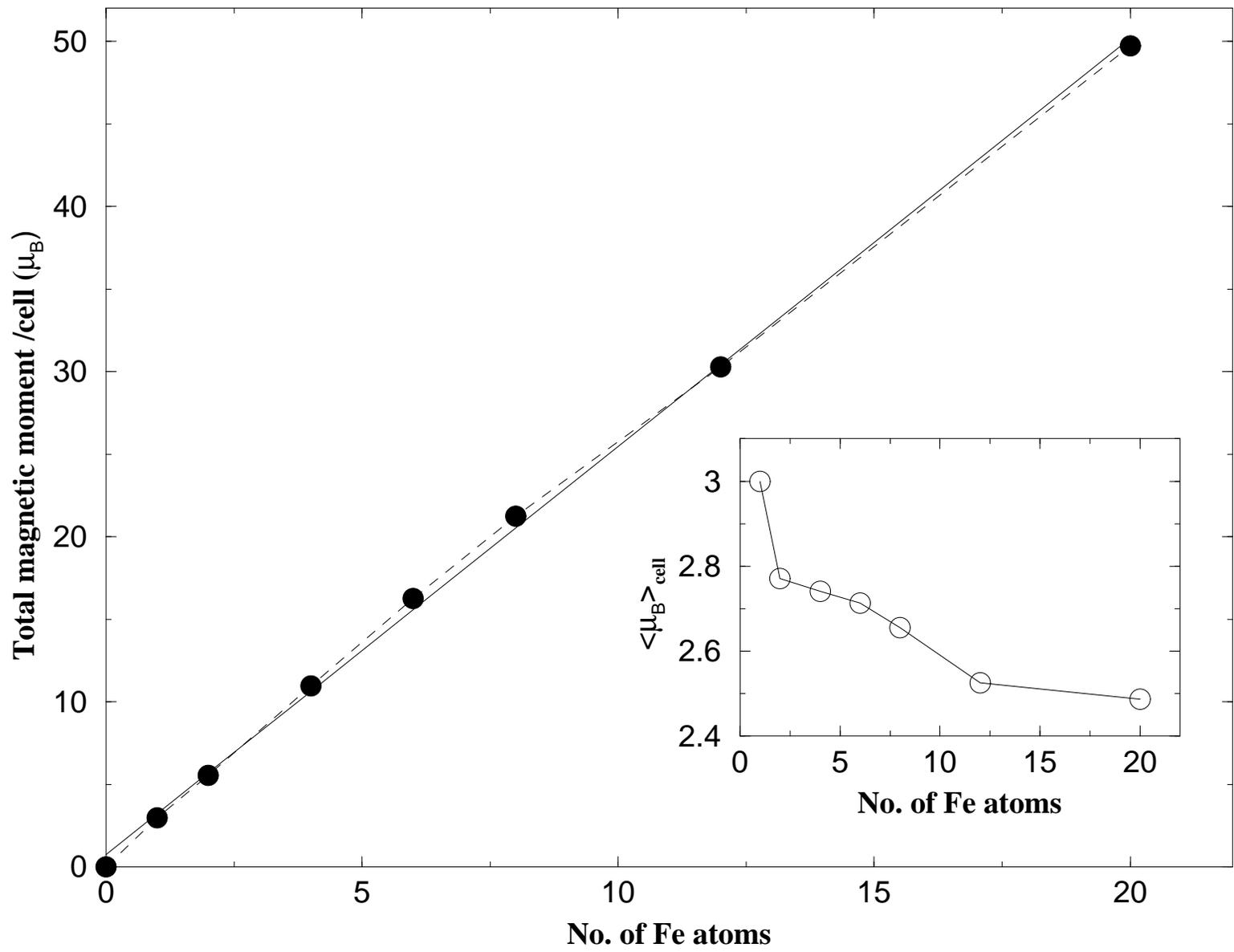
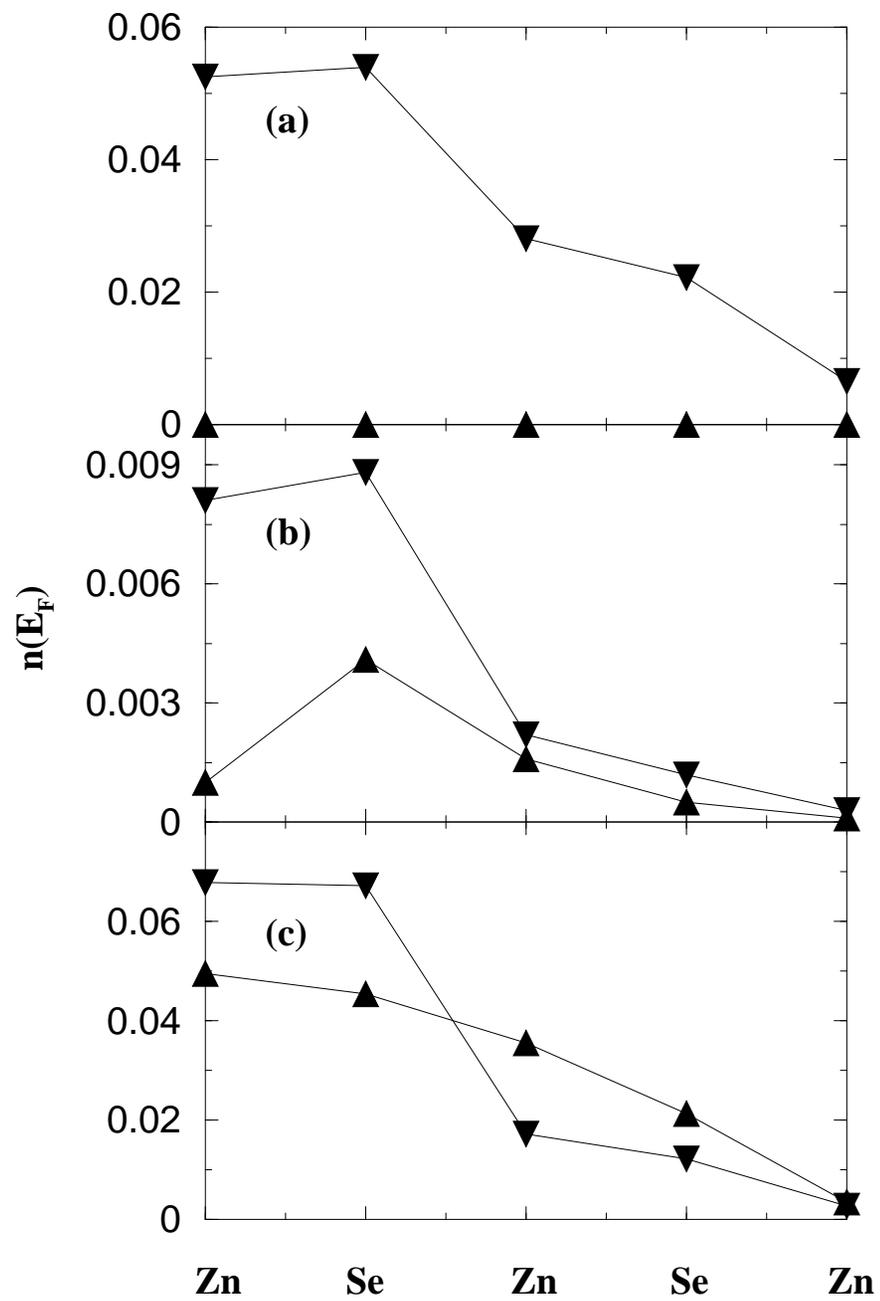
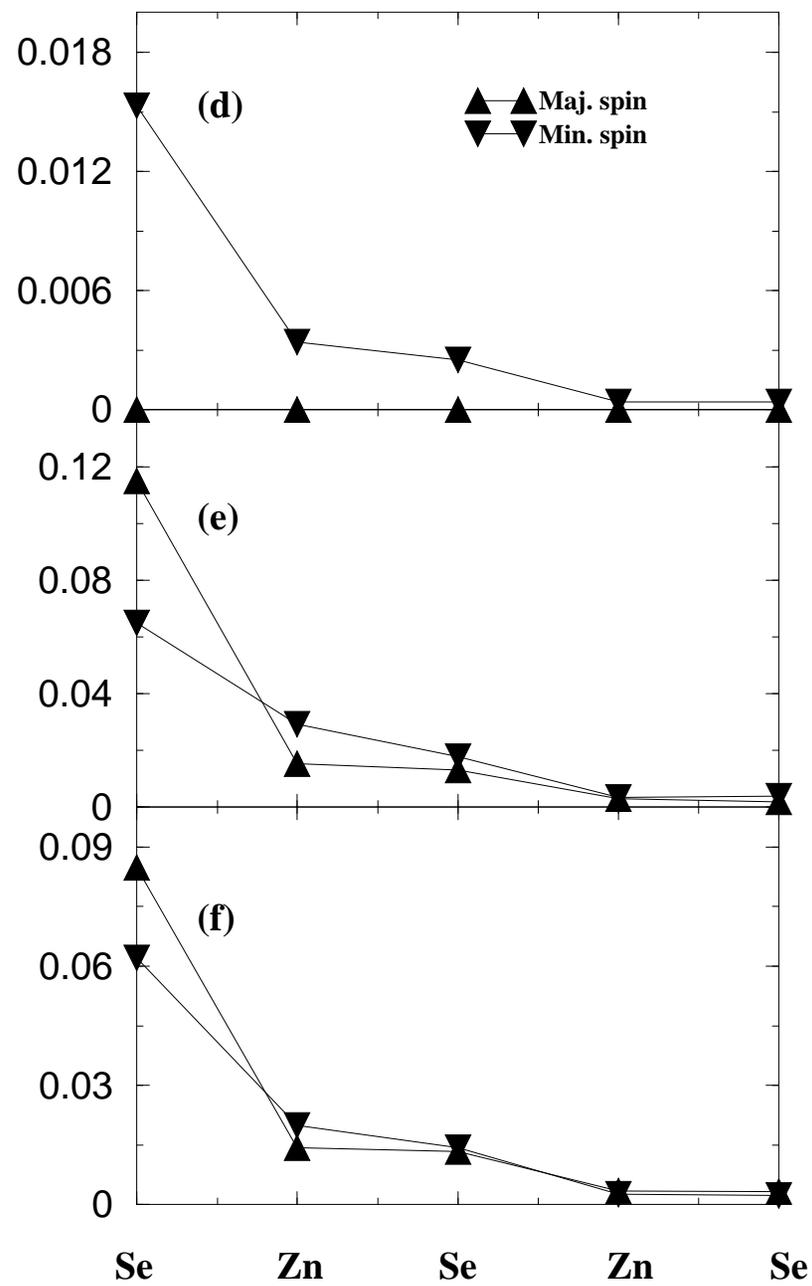
\end{document}